\PassOptionsToPackage{dvipsnames}{xcolor}
\documentclass[conference]{IEEEtran}
\IEEEoverridecommandlockouts
\usepackage{cite}
\usepackage{amsmath,amssymb,amsfonts}
\usepackage{algorithmic}
\usepackage{graphicx}
\usepackage{textcomp}
\usepackage{xcolor}
\usepackage{balance}
\usepackage{subcaption}
\usepackage[all]{nowidow}
\usepackage[hidelinks]{hyperref}
\usepackage[capitalise]{cleveref}

\newcommand{\name}{\textit{ElastiBench}}

\newcommand{\bettertextsim}{\raisebox{0.5ex}{\texttildelow}}

\begin{document}

\title{\name{}: Scalable Continuous Benchmarking on Cloud FaaS Platforms}

\author{\IEEEauthorblockN{Trever Schirmer, Tobias Pfandzelter, David Bermbach}
    \IEEEauthorblockA{\textit{Technische Universit\"at Berlin \& Einstein Center Digital Future}\\
        \textit{Scalable Software Systems Research Group} \\
        \{ts,tp,db\}@3s.tu-berlin.de}
}

\maketitle

\begin{abstract}
    Running microbenchmark suites often and early in the development process enables developers to identify performance issues in their application.
    Microbenchmark suites of complex applications can comprise hundreds of individual benchmarks and take multiple hours to evaluate meaningfully, making running those benchmarks as part of CI/CD pipelines infeasible.
    In this paper, we reduce the total execution time of microbenchmark suites by leveraging the massive scalability and elasticity of FaaS (Function-as-a-Service) platforms.
    While using FaaS enables users to quickly scale up to thousands of parallel function instances to speed up microbenchmarking, the performance variation and low control over the underlying computing resources complicate reliable benchmarking.
    We present \name{}, an architecture for executing microbenchmark suites on cloud FaaS platforms, and evaluate it on code changes from an open-source time series database.
    Our evaluation shows that our prototype can produce reliable results (\bettertextsim{}95\% of performance changes accurately detected) in a quarter of the time ($\leq$15min vs.~\bettertextsim{}4h) and at lower cost (\$0.49 vs.~\$1.18) compared to cloud-based virtual machines.
\end{abstract}

\section{Introduction}
\label{sec:introduction}

Microbenchmarks offer a fine-grained insight into the performance of an application by repeatedly benchmarking a small part of the application, e.g., on the function or package level~\cite{Laaber_2021_predicting_unstable}.
This allows developers to gain detailed insight into the impact their code changes have on specific code paths.
To achieve this, microbenchmarks can be run as part of a CI/CD (continuous integration and deployment) pipeline, which lets developers detect and fix performance regressions before they are introduced to the live environment~\cite{BencherDev, Anzt_2019_ContBench, Pearce_2023_CollabBench}.

Current approaches of executing microbenchmarks in cloud environments rely on virtual machines, which requires multiple repetitions of every microbenchmark to achieve reliable and statistically significant results~\cite{grambow2022microbenchmark,Laaber_2019_mb_cloud,book_bermbach2017_cloud_service_benchmarking}.
For example, in previous work on microbenchmarking in open-source code bases, we needed to repeat benchmarks more than 40 times on different virtual machines to achieve reliable results.
This took \bettertextsim{}4h and cost \bettertextsim{}\$1.14 for \emph{VictoriaMetrics},\!\!\footnote{\url{https://github.com/VictoriaMetrics/VictoriaMetrics}} and took \bettertextsim{}11h and cost \bettertextsim{}\$3.15 for \emph{InfluxDB}\footnote{\url{https://github.com/influxdata/influxdb}}~\cite{grambow2022microbenchmark}.
These benchmark durations and costs are prohibitive for integrating performance regression detection in CI/CD pipelines for \emph{every} release (let alone every commit), where continuous evaluation with rapid developer feedback is paramount.

\begin{figure}
    \centering
    \includegraphics[width=0.5\textwidth]{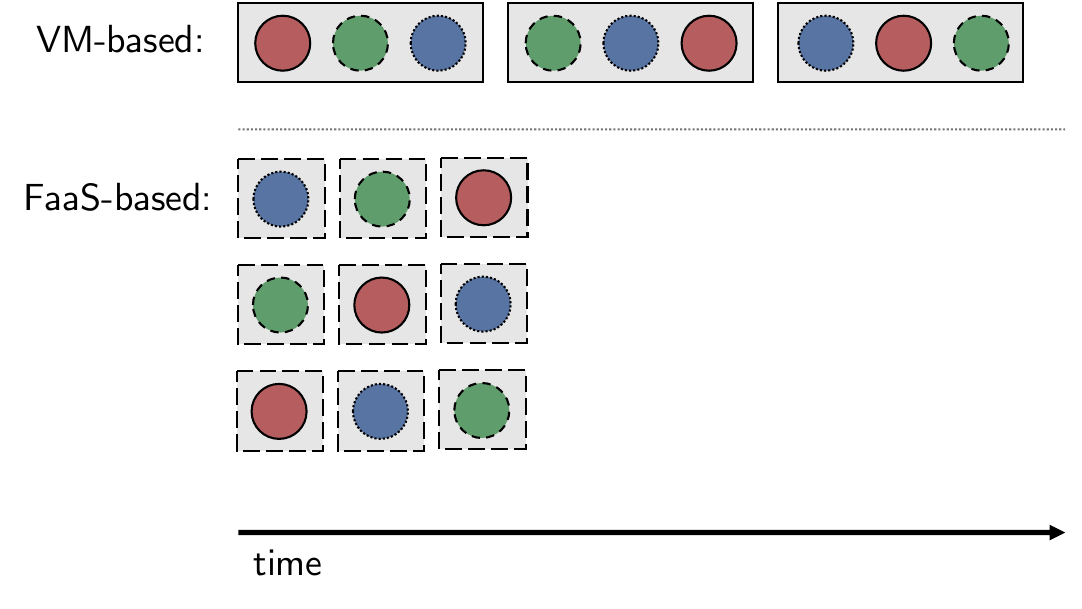}
    \caption{The traditional approach of microbenchmarking (top) relies on executing tasks (different circles) in random order multiple times on different virtual machines (gray boxes) to get reliable results. Using FaaS, these tasks can be executed on multiple function instances in parallel (bottom). With instance parallelism (three in this example), the duration of the suite run can be drastically reduced while also reducing inter-microbenchmark influences.}
    \label{img:overview}
\end{figure}

We propose leveraging Function-as-a-Service (FaaS) to massively scale out microbenchmark execution with little lead time, as shown in \cref{img:overview}:
The elastic scalability of cloud FaaS platforms allows us to run hundreds of microbenchmarks in parallel, achieving robust results in a fraction of the time sequential microbenchmark execution requires.
This enables developers to run them often as a normal part of their development workflows, e.g., to evaluate whether a local change has performance implications.

The key challenge in deploying microbenchmarks on FaaS is the unique execution environment that is prone to performance variability through cold-starts, temporal variability, and infrastructure heterogeneity~\cite{paper_bermbach2020_faas_coldstarts,schirmer2023nightshift}.
With \name{}, we overcome this by focusing on the detection of relative performance regression between two versions of the same software:
For this, we deploy both versions in the same FaaS function instance (similar to duet benchmarking~\cite{bulej2020duet}), thus ensuring an identical execution environment.
We then repeatedly call this function, triggering multiple microbenchmarks runs (see also \cref{img:overview}).
By carefully managing the execution order of microbenchmarks and their targets as well as the statistical analysis process afterwards, we can mitigate the noise effects of the FaaS platform and maintain the robust detection capabilities of the microbenchmark suite.
At the same time, we gain the significant speed-up and cost reductions of the FaaS environment, making our approach far superior to the state-of-the-art.
In this regard, we present the following contributions:

\begin{itemize}
    \item We analyze the main challenges in microbenchmarking using FaaS~(\cref{sec:mb-faas}).
    \item We propose \name{}, an approach for executing microbenchmark suites on cloud-based FaaS platforms~(\cref{sec:arch}).
    \item We implement a proof-of-concept prototype of \name{} for Go microbenchmarks on AWS Lambda (\cref{sec:impl}).
    \item We evaluate the detection capabilities and the efficiency of \name{} by comparing it to published results that used cloud VMs~(\cref{sec:eval}).
\end{itemize}

\section{Microbenchmarking in the Cloud}
\label{sec:microbenchmarking}

Microbenchmarks are short-running performance tests that evaluate a Software Under Test (SUT) on a smaller scale than full application benchmarks~\cite{grambow2022microbenchmark}.
Their scope can be compared to that of unit tests, i.e., they measure the performance of a single method or package~\cite{Laaber_2019_mb_cloud}.\!\!\footnote{To disambiguate from FaaS functions, we refer to the component that is being evaluated as methods, independently of what they are called in their respective programming language.}
Previous research has shown that microbenchmarks are easier to set up than application benchmarks, which can require other components to run~\cite{paper_grambow2021_optimizing_microbenchmarks}.
In compiled languages, the corresponding microbenchmark test files are compiled and then repeatedly executed.
In the Go programming language, for example, every method that is in a file with the suffix \texttt{\_test.go} and starts with \texttt{Benchmark} counts as a microbenchmark and can be executed using the \texttt{go test} command line tool~\cite{Go_2024_testing}.
With a configurable overall duration, the tool then reports the total amount of iterations that were executed, and the average time per execution.
Usually, executing a single benchmark takes on the order of a few seconds~\cite{Laaber_2019_mb_cloud,Cheney2019highperformance}.

Especially in cloud environments using shared resources, the total duration of a microbenchmark varies between different runs.
This is due to the inherent variability of the code that is executed, and effects within as well as between instances~\cite{grambow2022microbenchmark}.
When trying to measure the application performance, the goal is to minimize the impact of instance variability to capture the true performance of the benchmark.
Thus, the goal is not to achieve a consistent result, as the microbenchmark itself might be highly variable, especially if it is written in an interpreted language~\cite{Laaber_2019_mb_cloud}.
To limit measurement error, the order in which microbenchmarks are executed is randomized (Randomized Multiple Interleaved Trials, RMIT~\cite{abedi2017conducting}) so that order-effects are averaged out.
Next, to limit the impact of performance variance between instances, the two different versions of the SUT are executed on the same virtual machine and only their relative difference is taken into account for performance measurements.
These experiments are then repeated on different virtual machines~\cite{grambow2022microbenchmark}.
Current research indicates that 5 to 30 iterations of a microbenchmark are enough to get reliable results~\cite{Laaber_2019_mb_cloud}.

Developers use microbenchmarks to detect whether a code change had performance implications.
Thus, microbenchmarking results from the previous and new version of the software are collected and analyzed to detect statistically significant changes.
Current research uses the median performance change to measure the difference between two versions of a microbenchmark because it is more robust to outliers and skewed data than averages~\cite{Maricq_2018_Taming}.
A standard methodology to calculate the confidence interval of the median is bootstrapping, which we also used in our previous work~\cite{kalibera2020bootstrapping, grambow2022microbenchmark,Bulej_2017_bootstrapping,Bulej_2017_bootstrapping_b,scipy_bootstrap,Maricq_2018_Taming}.
Bootstrapping randomly resamples from the population with replacement to generate multiple bootstrap samples and calculates their median.
These random samples are then used to calculate confidence intervals around the measured median.
According to Laaber et al.~\cite{Laaber_2019_mb_cloud}, the underlying performance variability in microbenchmarks executed in the cloud is so high that a small performance change (3\% to 10\% in different studies\cite{Georges_2007_PerformanceChange3,Mytkowicz_2009_PerformanceChange10}) does not necessarily imply a real-world change in performance even if it is statistically significant.
\section{Challenges for Microbenchmarking on FaaS}
\label{sec:mb-faas}

FaaS platforms enable developers to easily create massively scalable applications.
The main use case for serverless platforms are short-running, resource-light functions that take on the order of seconds to complete~\cite{Eismann_2021_Review}.
FaaS platforms are an efficient execution environment for such tasks as they can benefit from bin-packing different small function executions from different tenants on shared hardware.
Cloud FaaS platforms in particular rely on optimized execution environments to quickly spin up function instances on demand, i.e., cold starts, and by overprovisioning resources~\cite{Manner_2018_Coldstarts,Agache_2020_Firecracker,Brooker_2023_AWSonDemand}.
This has interesting implications for using FaaS platforms to run microbenchmarks, which are sensitive to performance variability given their short execution duration.

\subsection{Performance}

Due to their small resource footprint, the FaaS provider can run many function instances on the same computing resources.
The smallest functions on Google Cloud Functions and AWS Lambda, for example, have access to less than 10\% of one vCPU~\cite{Cordingly_2022_CPUTams}, which leads to high interference between functions sharing compute resources.
Previous research has shown that function performance can vary up to 15\% diurnally and, even if executed at the same time, the variation between function instances can be considerable~\cite{schirmer2023nightshift}.
These performance variations increase the uncertainty of FaaS-based microbenchmarking results, as it is unknown to which degree an experiment has been affected by underlying performance changes.
Additionally, the performance difference between a cold and warm start can also influence benchmarking results, as during a cold start the application code might not be optimized yet, and some parts of the file system might be lazily loaded~\cite{Brooker_2023_AWSonDemand}.

\subsection{Restricted Environment}

To enable running isolated function with minimal overhead, the execution environment of a FaaS function is more restricted than a virtual machine.
The file system, for example, might be read-only except for special directories, and operating system features such as mounting file systems or (nested) virtualization might be missing.
When writing FaaS applications, it is possible take platform-specific limitations into account.
However, microbenchmarking suites might not be designed to run in restricted environments, which can lead to invalid results or test failures.
Thus, it is advisable to adapt microbenchmarking suites to work in these restricted environments if developers want to execute as many microbenchmarks as possible.

\subsection{Representative Environment}

When executing microbenchmarks, the experiment environment should resemble the real use case as much as possible to achieve relevant results~\cite{book_bermbach2017_cloud_service_benchmarking}.
FaaS platforms offer a standardized interface with only a few configuration options, which might be limiting for use cases with unique resource demands.
Additionally, while the platforms allow developers to choose a processor architecture, the exact CPU model used is not configurable and has been shown to vary within individual availability zones of a provider~\cite{Cordingly_2023_Sky}.
\section{Executing Microbenchmarks in FaaS}
\label{sec:arch}

\begin{figure*}
    \centering
    \includegraphics[width=\textwidth]{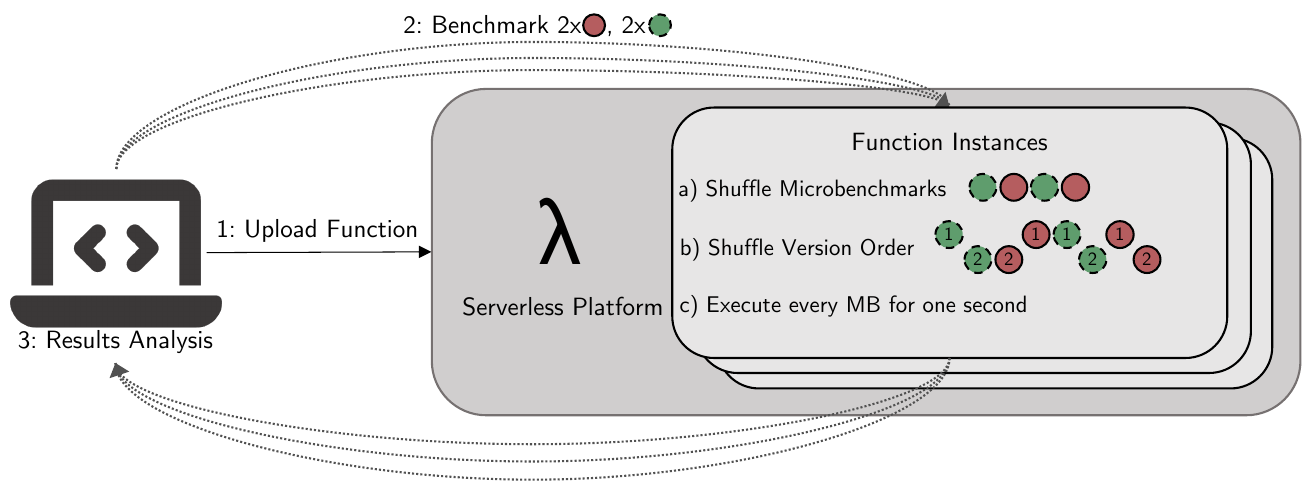}
    \caption{The process used to collect benchmarking data from FaaS functions. First, the function image (cf.~\cref{sec:impl}) is built and deployed to the FaaS platform. In step two, the function is called repeatedly with configurable repeats per microbenchmarks and instance parallelism. The results of all calls are then analyzed on the calling machine. The calling system can be the workstation of a developer or an automated CI/CD pipeline. While this figure shows multiple microbenchmarks being executed in one function call, \cref{img:overview} shows the extreme example of just one microbenchmark per function call.}
    \label{img:process}
\end{figure*}

We present \name{}, a system to conduct microbenchmarking experiments on FaaS.
\name{} can be used to conduct different kinds of microbenchmark experiments with a configurable function instance parallelism.

Execution of microbenchmarks is an embarrassingly parallelizable workload, since they have no dependencies on each other.
This makes them well suited as an application of FaaS.
Furthermore, two challenges of traditional microbenchmarking that are not as relevant on FaaS are execution-order effects, which might falsify results when microbenchmarks are executed in sequence on a single node, and the performance variability between a low number of virtual machine instances~\cite{grambow2022microbenchmark,Laaber_2019_mb_cloud}.
When using FaaS functions with high parallelism, the impact of a function instance with outlier performance is lower due to the sheer number of instances.
Additionally, the execution order of benchmarks within a function instance can be randomized as well: by only executing one microbenchmark per function invocation, the platform assigns every microbenchmark to an instance using its own control logic, which means that the order effects are averaged out with sufficiently many repeats.

Otherwise, the same limitations that apply when running microbenchmarks on cloud-based virtual machines~\cite{grambow2022microbenchmark,Laaber_2019_mb_cloud} apply to FaaS as well:
Due to the inherent performance variability of cloud platforms, a microbenchmark should always be run for each version of the software within the same function instance.
In the statistical analysis, only the relative performance change between these versions is taken into account.
This relative change is transitive, i.e., subsequent versions of the software only need to be compared to a reference version and their relative difference to other versions can be calculated based on that.

\cref{img:process} shows the process for running microbenchmarks on a FaaS.
The runner that is used to conduct performance benchmarks (e.g., the workstation of a developer or a CI/CD runner) first builds the function image that will be called.
The image contains the two software versions and the code required to run microbenchmarks (cf. \cref{sec:impl}).
Afterwards, the newly deployed function is called by the runner with the following parameters: the microbenchmarks that should be executed, whether their order and the order in which to run the different SUT versions should be randomized, and timeouts for the benchmark runs.
The microbenchmark execution can be repeated many times in different function calls.
This way, the system takes advantage of the transparent assignment of calls to function instances: by randomizing the order of function calls, the instances on which the repeats of microbenchmarks are executed is also effectively randomized.
After all microbenchmarks have been executed sufficiently often, the function is obsolete and can be deleted: since the SUT source is part of the function image and the comparison of these two versions has been finished, the function will not be called again.
All other container layers of the function image (e.g., the actual function) can be reused for future experiments.
Depending on the number of microbenchmarks and the instance parallelism, the whole process can complete within the maximum time the FaaS platform keeps a function instance alive (e.g., \bettertextsim{}15min for AWS Lambda and Google Cloud Functions).
Since the short overall duration of a benchmark run essentially means that only the first requests will be cold starts, the runner can influence the total duration and total cost of the benchmarking run by configuring the amount of requests that are sent in parallel: higher parallelism leads to shorter runs, while increasing cost due to the increased number of cold starts.

\section{Proof-of-Concept Implementation}
\label{sec:impl}

In this section, we present the prototype function we implemented to execute microbenchmarks with the \name{} approach.
We focus on SUTs written in Go, as it is a popular programming language for cloud systems and has native support for running microbenchmarks.
We implement our prototype for AWS Lambda since it is one of the most mature and popular platforms and showed comparably low performance variation in previous research~\cite{schirmer2023nightshift,Copik_2021_Sebs}.
The prototype is available as open-source\footnote{\url{https://github.com/umbrellerde/microbenchmarks-faast}}\!\!.
The architecture and prototype do not require any features that are specific to AWS Lambda, so that it could easily be adapted for other platforms.
We describe the components comprising the function image in the following sections. The total file size of an image for a complex SUT can be bigger than \bettertextsim{}1GB, of which \bettertextsim{}240MB are components that are necessary for the microbenchmarks to work, and the rest is taken up by the SUT as well as a prepopulated Go build cache of the SUT source code.

\subsubsection*{Software Under Test}

The image contains two folders (\texttt{v1} and \texttt{v2}) with the source code of the two software versions that should be compared.
Developers could reduce the size of these folders by excluding files that are not necessary for benchmarking (e.g., documentation, examples, ...) if they want to reduce cost and latency. For the project used in our evaluation, the unoptimized total size is \bettertextsim{}240MB.

\subsubsection*{Microbenchmarking Pipeline}

This component compiles the SUT and runs microbenchmarks using the Go tool chain. The total size of the tool chain in our prototype is \bettertextsim{}230MB.

\subsubsection*{Benchrunner}

The \emph{Benchrunner} is the component that is registered as entry point for function calls from the platform and runs the microbenchmarking pipeline in a configurable way (e.g., randomizing the order in which benchmarks are called and configuring timeouts). The total file size in our prototype is \bettertextsim{}7MB.

\subsubsection*{Prepopulated Cache}

To speed up repeated compilation in benchmarks, the microbenchmarking pipeline uses a build cache that stores precompiled binaries.
Without a prepopulated cache, cold starts would take even longer as the cache would have to be filled during the first call.
Thus, we fill this cache during the build process (i.e., on the machine of the developer).
While this increase in image size (almost 1GB for the project used in our evaluation) also increases cold start times since more data needs to be transferred to function runners, it still leads to overall faster and more consistent performance.
Due to caching in function runners~\cite{Brooker_2023_AWSonDemand}, the first cold starts after deploying a new version of the function are slower, but subsequent cold starts benefit from caching of these larger images.

\subsubsection*{Instance Cache}

Due to the way that FaaS providers limit writable locations in the file system (see also \cref{sec:mb-faas}), the prepopulated cache location is not writable during function execution.
To work around this, Go supports custom cacher implementations~\cite{gocacheprog}.
In the function image, a custom cacher is used to read the prepopulated cache but writes changes to another (writable) directory.
When the microbenchmarking pipeline tries to access cached binaries, the custom cacher first checks the instance cache and then falls back to the prepopulated cache. The file size of the custom cacher program is \bettertextsim{}3MB.
\section{Evaluation}
\label{sec:eval}

In this section, we present how the experiments are designed (\cref{subsec:all-exp-design}), their results (\cref{subsec:all-exp-results}), and discuss key takeaways (\cref{subsec:all-exp-takeaways}).

\subsection{Experiment Design}
\label{subsec:all-exp-design}

The following sections contain descriptions of the SUT used in our evaluation (\cref{subsec:eval-sut}), the specific methods we use for our analysis (\cref{subsec:eval-stats}), and give an overview of the experiments (\cref{subsec:eval-experiments-overview}).

\subsection*{Software Under Test}
\label{subsec:eval-sut}

Our evaluation focuses on the open-source time-series database software VictoriaMetrics, as we have used it in previous work and have reliable performance results to compare \name{} to~\cite{grambow2022microbenchmark}.
We repeat the same microbenchmarks with the same SUT versions that were used in the original paper and compare their results.

\subsection*{Statistical Analysis}
\label{subsec:eval-stats}
\begin{figure}
    \centering
    \includegraphics[width=\linewidth]{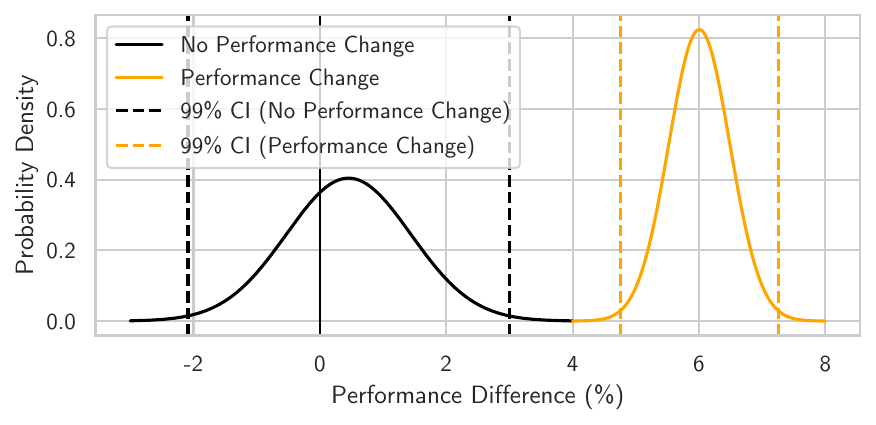}
    \caption{When comparing the difference in performance between two versions, the confidence interval shows range where the real difference likely is. If the CI overlaps zero (black curve), we detect no performance change. If it does not overlap zero (yellow curve), we detect a \textit{performance change}.}
    \label{img:demo_significant}
\end{figure}

For this evaluation, all microbenchmarks which collect less than 10 results are ignored.
Microbenchmarks might fail because they cannot run (e.g., a build failure), or because they ran for more than twenty seconds, after which they are interrupted.
This can happen for some benchmarks that require extensive setups or have long iterations.
We argue that these are not relevant as single microbenchmark executions microbenchmarks longer than one second with the default Go benchmark parameters are indicative of faulty benchmark implementations~\cite{Cheney2019highperformance}.
In an experiment, multiple microbenchmarks are run.
Some microbenchmark executions are run for different configurations, e.g., different input sizes for a function.
We treat these configurations as independent microbenchmarks.
If the 99\% confidence interval (CI) of the median difference in performance between the two SUT versions that we calculated using bootstrapping (see \cref{sec:microbenchmarking}) does not overlap 0, then the experiment has detected a \textit{performance change} for that microbenchmark.
If one experiment detects a \textit{performance change} and another one does not, we call this a \textit{possible performance change}, as it was not reliably detected across multiple experiments.
\cref{img:demo_significant} (note the x-axis values) shows an example of a distribution that has a \textit{performance change} as opposed to \textit{no performance change}.

When comparing the results of two experiments with each other, we define them as \textit{agreeing} with each other if both find a \textit{performance change} in the same direction, or if both find \textit{no change} (i.e., both confidence intervals overlap 0).
Otherwise, they \textit{disagree}.
Note that two experiments \textit{agreeing} on a \textit{performance change} does not take the magnitude of the change into account, only the sign of the effect.
To measure how close the magnitude of two experiments that find a \textit{performance change} is, we calculate the relative number of microbenchmarks for which the reported median of the first experiment is inside the CI found by the other experiment (\textit{one-sided coverage}) and the relative number of microbenchmarks for which both results are within the CI reported by the other experiment (\textit{two-sided coverage}).

\subsection*{Experiment Overview}
\label{subsec:eval-experiments-overview}

Unless otherwise noted, we deploy all functions with a timeout of $15$min, which is the maximum timeout allowed by AWS Lambda.
The functions have 2048 MB RAM, and access to 1.29 vCPUs using an ARM architecture.
Every function call contains one microbenchmark call, which is repeated three times inside the function.
This call is repeated 15 times per microbenchmark, leading to 45 results for every microbenchmark.
This is the same number used in our previous study~\cite{grambow2022microbenchmark} and above the number of repeats found in other research~\cite{Laaber_2019_mb_cloud}.
We limit the call parallelism of the call script to 150, i.e., a maximum of 150 microbenchmarks execute on the FaaS platform in parallel.

Every microbenchmark is repeated until a stable number of executions per second is found.
The SUT in these experiments is the open-source time series database VictoriaMetrics.
The two software versions we compare are the initial (\texttt{f611434}) and last commit (\texttt{7ecaa2fe}) evaluated in our previous work~\cite{grambow2022microbenchmark}, which we refer to as the \emph{original dataset} in this paper.
In this paper, we run the following experiments:

\begin{enumerate}
    \item \textbf{A/A Experiment:} In the A/A experiment, both versions of the SUT are the initial commit used in our previous work~\cite{grambow2022microbenchmark}.
          With this experiment, we verify that the system works and that the variability of the FaaS platform does not invalidate the results.
    \item \textbf{Baseline Comparison Experiment:} In the baseline experiment, we execute exactly the configuration described above.
    \item \textbf{Replication Experiment:} We then repeat this experiment again to measure the consistency of \name{} between runs.
    \item \textbf{Lower Memory Experiment:} In this experiment, we reduce the amount of memory available to functions to 1024 MB.
          This allows us to measure the impact of platform configuration on the robustness of \name{}.
    \item \textbf{Single-Repeat Experiment:} In this experiment, we reduce the number of microbenchmark repetitions inside the function from three to one, and increase the number of function calls for every microbenchmark from 15 to 45.
          This way, the overall number of results per microbenchmark stays the same (45).
          This experiment allows us to measure the impact multiple repeats of the same microbenchmark in the same function call have.
    \item \textbf{Repeats Necessary for Consistent CI Size:} In this experiment, we measure how many repetitions are necessary to achieve the same or a smaller CI size than the original dataset.
          We repeat the baseline experiment with 45 function call repeats instead of 15, leading to 135 measurement results per microbenchmark.
          We then repeatedly calculate median difference in performance and its CI with a growing number of results.
          For all results for which the ultimate CI overlaps with the CI in the original dataset (i.e., they have a common value), we measure how many repeats are required to have a CI size that is less than or equal to the CI size in the original dataset.
\end{enumerate}

\subsection{Experiment Results}
\label{subsec:all-exp-results}

We first report the results of the all experiments that collect the same data as the original dataset (\cref{subsec:exp-aa,subsec:exp-baseline,subsec:exp-mem,subsec:exp-repeat,subsec:exp-single}), and then compare their results against each other (\cref{subsec:uncertain}). Afterwards, we analyze how many repeats are necessary to achieve a consistent CI size (\cref{subsec:exp-repeats-necessary}).

\subsubsection{A/A Experiment}
\label{subsec:exp-aa}

\begin{figure}
    \centering
    \includegraphics[width=\linewidth]{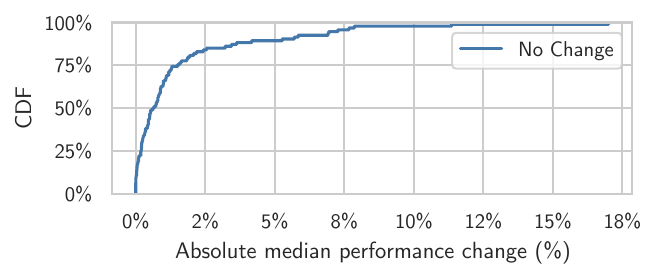}
    \caption{CDF showing the performance differences identified in the A/A experiment. While some microbenchmarks have a high difference, all of them are correctly categorized as \textit{no performance change}.}
    \label{img:effect_size_aa}
\end{figure}

Out of 106 microbenchmarks in the SUT, 90 were executed successfully.
Out of those, no \textit{performance change} was found.
This shows that \name{} does not mistakenly detect \textit{performance changes} in unchanged software.
\Cref{img:effect_size_aa} shows an overview of the performance differences found.
The median difference in performance is 0.047\%, with a maximum of 32\%.
Large performance differences indicate that a microbenchmark is highly variable (as the same version is executed twice), which is correctly detected as \textit{no performance change}.
Running the experiment took \bettertextsim{}8min and cost \$1.18\footnote{It was started \bettertextsim{}2024-05-12T17:35:00 UTC}\!\!.

\begin{figure}
    \centering
    \includegraphics[width=\linewidth]{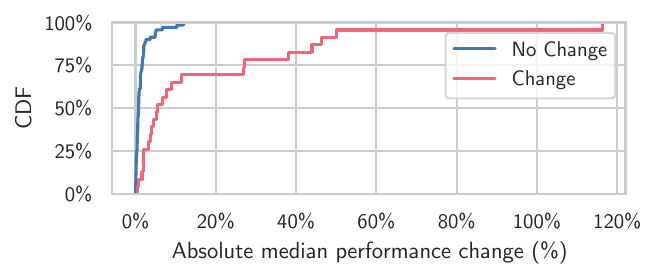}
    \caption{CDF showing the performance differences identified in the baseline experiment. \textit{Performance changes} have a generally higher difference than non-changes, with the median \textit{performance change} being 3.08\%.}
    \label{img:effect_size_baseline}
\end{figure}

\subsubsection{Baseline Experiment}
\label{subsec:exp-baseline}

We show an overview of the differences in performance measured in the baseline experiment in \cref{img:effect_size_baseline}.
The median \textit{performance change} of the two SUT versions is 4.71\%.
The maximum difference in performance of a \textit{change} and \textit{no change} are 116\% and 26\% respectively.

After removing microbenchmarks for which only one experiment contains results, there are 91 (out of 106) microbenchmarks that can be compared to the original dataset.
Out of those, both datasets \textit{agree} for 95.65\% of microbenchmarks (88 overall).
The disagreements are as follows: for 3.2\% of microbenchmarks (three overall), both find a performance change but identify different directions.
These three results are from a single microbenchmark function (\texttt{BenchmarkAddMulti}) with different configurations.
In every case, the original dataset identifies a \textit{performance change} of \bettertextsim{}-10\%, while the new experiments identifies a positive 5-7\% \textit{performance change}.
This benchmark was changed between the two SUT versions\footnote{commit hash eb103e15} with a considerable change in the benchmark itself, which leads to inconsistent results as the two versions are essentially different benchmarks.
We argue that this is not representative for microbenchmarks, so that the difference in results between the original dataset and our experiments are not indicative of failure of \name{}.

The remaining \textit{disagreement} is a single microbenchmark for which the original dataset does not find a performance change while the new dataset finds one.
The difference in performance is 1.96\% in the original dataset, and 0.60\% in the new dataset.
As previous research has shown that a performance difference this small is not reliably detectable with cloud microbenchmarks (cf.~\cref{sec:microbenchmarking}), we argue that while this is technically a disagreement, it is an issue of cloud benchmarking in general.

The median \textit{performance change} measured in this paper has a \textit{one-sided coverage} with the original dataset for 86.96\% of microbenchmarks, and 52.17\% vice versa.
The \textit{two-sided coverage} is 50\% of microbenchmarks, i.e., for 50\% of \textit{performance changes} in a microbenchmark, the result of one experiment is within the CI of the other result.
We argue that the overall high \textit{agreement} of the baseline experiment with the original dataset demonstrates that both find the same \textit{performance changes}.
The magnitude of the performance difference however depends on many factors that were changed between the baseline experiment and our original work\cite{grambow2022microbenchmark}, e.g., the execution environment and version of the programming language, which explains the comparatively low \textit{two-sided coverage}.
The experiment took \bettertextsim{}11min, and the total cost for the experiment is \bettertextsim{}0.18\$, i.e., a similar cost to the original experiments while taking \bettertextsim{}4.6\% of time\footnote{It was started \bettertextsim{}2024-05-12T16:50:00 UTC}\!\!.

\subsubsection{Replication Experiment}
\label{subsec:exp-repeat}
The replication experiment has the same \textit{agreement} with the original dataset as the baseline dataset.
The \textit{one-sided coverage} with the original dataset is 81.72\%, and 51.61\% in the other direction.
The \textit{two-sided coverage} is just 48.39\%
It \textit{disagrees} with the baseline experiment for 10.87\% of microbenchmarks, which are all microbenchmarks that are not in the original dataset (i.e., they were not run successfully or had too few runs to be included).
The maximum difference in performance found by one experiment where the other experiment did not find a \textit{performance change} (the \textit{possible performance change}) was 5.25\%.
Overall, the results show that while the changes with a higher performance difference are detected continuously, smaller changes are not detected robustly.
This maximum \textit{possible performance change} of 5.25\% is still within the limits found in other work (cf. \cref{sec:microbenchmarking}).
The experiment cost \$1.18 to execute and took \bettertextsim{}9min\footnote{It was started \bettertextsim{}2024-05-12T19:35:00 UTC}\!\!.

\subsubsection{Lower Memory Experiment}
\label{subsec:exp-mem}
The maximum amount of memory used by a microbenchmark in the previous experiments was 740 MB, so that we reduce the memory footprint of the functions to 1024 MB for this experiment.
This reduces the availability memory reduces compute resources to 0.255vCPUs.
The lower memory experiment \textit{agrees} with the original dataset the same way both previous experiments do.
Overall, only 81 microbenchmarks were successfully executed, i.e., some microbenchmarks did timeout due to the reduced compute capacity.
For ten microbenchmarks (\bettertextsim{}20\%), this experiment \textit{disagrees} with the baseline experiment, with a maximum \textit{possible performance change} of 6.22\%.
The experiment cost \$0.69 to execute and took \bettertextsim{}12min\footnote{It was started \bettertextsim{}2024-05-12T19:10:00 UTC}\!\!\!\!.

\subsubsection{Single Repeat Experiment}
\label{subsec:exp-single}
In the single repeat experiment, we again find the same \textit{agreement} with the original dataset as in the experiments above.
The single repeat experiment \textit{disagrees} with the baseline experiment for 18 microbenchmarks (\bettertextsim{}20\%) with a maximum \textit{possible performance change} of 5.09\%.
The experiment cost \$0.49 and took \bettertextsim{}17min\footnote{It was started \bettertextsim{}2024-05-12T20:40:40 UTC}\!\!\!\!.
Overall, repeating every microbenchmark once instead of three times leads to reduced overall cost while only minimally influencing results.

\subsubsection{Possible Performance Changes}
\label{subsec:uncertain}
\begin{figure}
    \centering
    \includegraphics[width=0.5\textwidth]{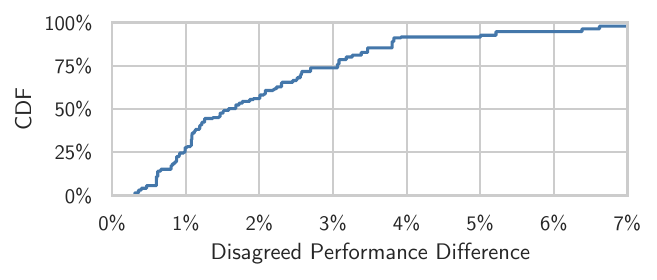}
    \caption{The maximum difference in performance of a microbenchmark when two experiments disagree whether a \textit{performance change} has happened.}
    \label{img:disagree_effect_size}
\end{figure}
We compare the difference in performance of all microbenchmarks where the baseline, replication, lower memory, and single-repeat experiments \textit{disagree} with each other and collect the maximum performance change that was found by one experiment.
An overview of these differences can be found in \cref{img:disagree_effect_size}.
The median difference is 1.58\%, the 75\textsuperscript{th} percentile is 3.06\%, and the biggest difference is 7.6\% in the \texttt{BenchmarkAdd/items\_100000} microbenchmark (which is unreliable, cf. \cref{subsec:exp-baseline}) between the single-repeat and lower memory experiments.

\subsubsection{Repetitions Necessary for Consistent CI Size}
\label{subsec:exp-repeats-necessary}
\begin{figure}
    \includegraphics[width=\linewidth]{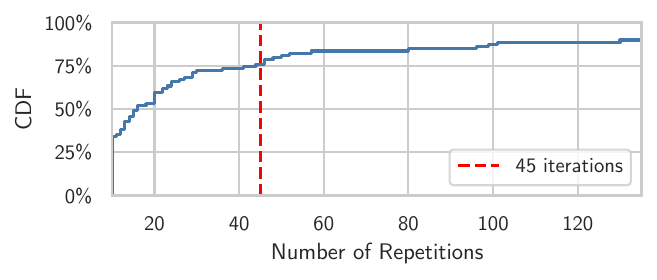}
    \caption{If the CI of the original dataset and the experiment overlap (i.e., they show a similar performance), this graph shows how many repetitions are necessary until the size of the CI of \name{} is equal to or less than the size the original dataset. The graph increases until 40 repetitions and then plateaus to around 90\% with more and more repetitions.}
    \label{img:many-repeats-leq}
\end{figure}

To evaluate how many repetitions of a single microbenchmark are necessary to achieve reliable results, we run an experiment where we repeat every microbenchmark 135 times: 3 times within a function call, which is called 45 times.
We then calculate the median and its CI repeatedly with more and more of the calls, which leads to increasingly accurate results.
\cref{img:many-repeats-leq} shows how close the results of \name{} can get to the original dataset.
Over all benchmarks where the CIs ultimately overlap each other (i.e., they share at least one common value), we calculate how many repetitions are necessary until the size of the CI is less than or equal to the CI size of the original dataset.
In 75.95\% of cases, this is achieved when using the same 45 repeats used in the original dataset.
At the full 135 steps, it is achieved for 89.87\% of microbenchmarks.

\subsection{Key Takeaways}
\label{subsec:all-exp-takeaways}

Our evaluation of \name{} shows that it is viable to execute microbenchmarking experiments on FaaS platforms, decreasing microbenchmark suite execution time to 6\% and lowering total execution costs compared to virtual machine-based experiments.
The different experiments show that the platform and experiment configuration seem to have minimal impact on performance changes with a sufficiently large difference that they can reliably be detected using cloud platforms.
For effect sizes of more than 7.06\% results stayed consistent between all experiment runs, with a median inconsistent change of 1.63\% (cf. \cref{subsec:uncertain}).
Other researchers have found that a relatively small significant performance change between 3\% and 10\% does not imply a real-world change due to the underlying variability in cloud platforms and microbenchmarks~\cite{Georges_2007_PerformanceChange3,Mytkowicz_2009_PerformanceChange10,Laaber_2019_mb_cloud}.
Our results fall within that range.

\newpage
\section{Discussion \& Future Work}
\label{sec:discuss}

Our evaluation shows how \name{} can lower execution time and cost of continuous microbenchmarking by leveraging the rapid elasticity and pay-per-use billing of cloud FaaS.
Nevertheless, we believe further optimizations are possible.

\subsection{Optimizing Function Configuration}

We use functions with 2048 MB of memory in our evaluation to ensure no microbenchmark runs out of memory, yet not every microbenchmark necessarily requires these resources.
Previous research has shown that it is possible to optimize FaaS function resource allocation further while the application is running~\cite{paper_schirmer2022_fusionize}.
By adapting the required resources per microbenchmark function, the total execution cost of the microbenchmark suite could be reduced further.
As most cloud FaaS platforms allocate compute resources with the configured memory size, however, care must be taken to not influence benchmark results, e.g., as using shared CPU cores could further increase performance variability.
Overall, however, there are limits to how low we can go in terms of resources as the SUT is, in the end, still a complete system such as VictoriaMetrics in our experiments.

\subsection{Benchmarking Strategy}

Our current benchmarking strategy with 45 repetitions of each microbenchmark (cf.~\cref{subsec:exp-repeats-necessary}) is sufficient to reduce the mean standard error of results that show a performance change to less than two percent, with an overall achievable standard error of around one percent.
In future research, we plan to further refine this strategy in order to increase the cost and time savings of \name{}.
We plan to evaluate the impact of repeating the microbenchmark within a function instance compared to repeatedly invoking a function.
Additionally, we will further assess the impact of cold start performance that results from calling multiple benchmarks in parallel.

\subsection{Microbenchmarks Beyond Go}

Our current prototype supports Go microbenchmarks, yet Go is not the only programming language with support for microbenchmarks:
Future extensions to our prototype could support Rust's \emph{cargo-bench}~\cite{rustcargobench}, \emph{Benchmark.js}~\cite{benchmarkjs}, Java's \textit{Microbenchmark Harness}\cite{JavaMH}, or the \emph{Google Benchmark Framework} for C++ projects~\cite{googlebenchmarkframework}.
Integrating such microbenchmark frameworks into \name{} will require further evaluation of their interaction with cloud FaaS performance variability.

\subsection{Limitations of the Lambda Runtime}

As described in \cref{sec:mb-faas}, the execution environment of cloud FaaS platforms is typically locked down to improve cold starts and multi-tenancy.
In our experiments of \name{}, a small fraction of microbenchmarks could not be executed successfully, e.g., because they write to the local file system.
On the one hand, these benchmarks could be repeated in a different environment without significantly increasing cost and duration of the entire microbenchmark suite.
On the other hand, a constrained execution environment challenges some assumptions benchmark authors make, e.g., why should the benchmark expect to be able to write to the local file system at all?
Running \name{} could provide an interesting avenue of such feedback for microbenchmark maintainers.

\section{Related Work}
\label{sec:related}

Quickly provisioning cloud resources is a convenient approach for running software performance benchmarks on powerful infrastructure without upfront investments~\cite{paper_grambow2019_continuous_benchmarking:_2019, Javed_2020_Perfci}.
Cloud variability, however, has necessitated special approaches to ensure reliable benchmarking results~\cite{Laaber_2019_mb_cloud,Leitner_2016_Patterns,abedi2017conducting}.

FaaS has proven to be useful for speeding up embarrassingly parallelizable problems in other contexts~\cite{werner2018serverless,Mileski_2022_Serverless_Parallel}, yet to the best of our knowledge \name{} is the first system to run general software microbenchmarks on FaaS platforms.
Previous research of benchmarking on FaaS platforms has focused on benchmarking the performance of the FaaS platform itself, e.g., in the form of functions, e.g.,~\cite{Eismann_2022_Stability, Lambion_2022_X86vsARM, Copik_2021_Sebs, Ginzburg_2021_Variability,Scheuner_2022_TriggerBench}, or entire FaaS applications, e.g.,~\cite{paper_grambow2021_befaas,grambow2023befaasextended}.

Further research on improving microbenchmark execution speed, cost, and reliability has focused on changing the duration of microbenchmarks, e.g., stopping experiments early~\cite{japke2023early, He_2019_Statistics, mittal2023adaptive}, or running only a selection of microbenchmarks without losing information on performance regressions~\cite{Laaber_2021_applying,Laaber_2022_multiObjectiveSearch,Mostafa_2017_Perfranker,De_2017_Perphecy,paper_grambow2021_optimizing_microbenchmarks}.
As these approaches could easily be integrated with the FaaS execution backend of \name{}, we consider them orthogonal to our approach.

\section{Conclusion}
\label{sec:fin}

In this paper we have presented \name{}, an approach for improving continuous microbenchmarking by leveraging the extensive elastic scalability of cloud FaaS platforms.
We identified challenges for running microbenchmarks using FaaS and presented an architecture that can execute microbenchmarks with similar robustness to the current state of the art.
Leveraging the massive parallelism of cloud FaaS, the system executes many microbenchmarks in parallel, with repeats on different function instances to minimize outside influences on results.
In our evaluation, our open-source prototype of \name{} on AWS Lambda executes microbenchmarks in less than 10\% of the time at the same or even lower cost than the current state of the art while achieving comparable results.
This enables developers to use microbenchmarks in more use cases where rapid feedback is important.

\section*{Acknowledgements}
Partially funded by the Bundesministerium f{\"u}r Bildung und Forschung (BMBF, German Federal Ministry of Education and Research) in the scope of the Software Campus 3.0 (Technische Universit\"at Berlin) program -- 01IS23068.

\newpage
\balance

\bibliographystyle{IEEEtran}
\bibliography{bibliography}

\end{document}